\documentstyle[A4,epsf,twocolumn]{article}
\begin{document}

\textwidth18.5cm
\textheight24.5cm
\topmargin-3cm
\oddsidemargin-1.0cm

\title{Testing operational phase concepts in quantum optics}

\author{ Jaroslav \v{R}eh\'{a}\v{c}ek$^{\dag}$, Zden\v{e}k
Hradil$^{\dag}$, Miloslav Du\v{s}ek$^{\dag}$,
Ond\v{r}ej Haderka$^{\ddag\dag}$, and Martin Hendrych$^{\ddag\dag}$\\
\dag \it Department of Optics, Palack\'{y} University,
17. listopadu 50, 772~00 Olomouc, Czech Republic\\
$\ddag$ \it Joint Laboratory of Optics of Palack\'{y}
University and Phys. Inst. Czech Acad. Sci., \\
\it 17. listopadu 50, 772~00 Olomouc, Czech Republic}
\date{}
 \maketitle

\begin{abstract}
An experimental comparison of several operational phase
concepts is presented. In particular, it is shown that
statistically motivated evaluation of experimental data
may lead to a significant improvement in phase fitting
upon the conventional Noh, Foug\`{e}res and Mandel
procedure. The analysis is extended to the asymptotic
limit of large intensities, where a strong evidence in favor
of multi--dimensional estimation procedures has been found.
\end{abstract}

\section{Introduction}
``The essence of quantum theory is its ability to predict
probabilities for the outcomes of tests, following specified
preparations'' \cite{peres98}. From a pragmatic point of view the
{\it quantum state\/} represents just our information on the system
corresponding to a particular preparation by a classical apparatus.
According to quantum theory this seems to be  the most complete
information.  However, the  accessibility of this information  is
questionable.
Not knowing the preparation procedure, one does not know the
quantum state of the system. There is no way to measure it for
a single realization of a quantum system.
The situation gets better if an ensemble of systems prepared in
the {\it same\/} quantum state are available. Then it is possible
to measure complementary observables in different experiments, and
the quantum state of the system can be inferred.
Since real ensembles are always finite, only the particular
numbers of occurrence of different results can be measured
instead of probabilities. This is a paradigm for an arbitrary
measurement. However, this scheme could seem purposless unless
theoretical predictions are compared with experiments. In quantum
domain this is not in general easy at all, and in practice many
sophisticated theories cannot be demonstrated on their experimental
counterparts.

The estimation of  phase differences in  in\-ter\-fe\-ro\-met\-ry
 appears to be a
nice example  of the above mentioned scheme, where the predictions
of quantum theory can be followed by an  experimental realization.
Optical measurements in  the domain of classical wave optics are
well established  and  belong to the most precise  measurement
schemes currently available. Significantly, such schemes may be
analysed in the framework of quantum phase.

 Quantization based on
the correspondence principle leads to the  formulation of
operational  quantum phase concepts \cite{noh92,noh93}. Further
generalization may be given in the framework of quantum estimation
theory; the prediction may be improved using the maximum
likelihood estimation. This improvement was
tested experimentally in matter wave optics with neutrons, and
a statistically significant improvement was
observed \cite{rehacek99b}. This is a
remarkable result, since the phase estimation is rather uncertain
for neutrons due to technical limitations of neutron
interferometry, where for example the visibility of interference
fringes is far below the ultimate value of $100 \%.$ In the
present paper the same theoretical background of optimal phase
estimation will be used for testing of phase resolution with
photons.

 Optical measurements offer many  advantages. Current optical
technology enables us to achieve  visibility of interference fringes
close to unity, and to very precisely set the intensities of light
pulses at levels deep below one photon on average.
As the main objective, different
strategies for accurate phase estimation  will be specified and their
consequences for achieved precision will be derived.

The paper is organized as follows. The mathematical tools
are reviewed in the next section, where the
operational phase concepts are naturally embedded in the
quantum estimation theory.
The experimental setup is described
in the third section.
A comparison of several phase estimation procedures based on
the experimentally measured data
is given in the fourth section.
Finally, the fifth section deals with the phase estimation
in the asymptotic regime.

\section{Phase estimation} \label{sec_theory}

The operational phase concepts can naturally be
embedded in the general scheme of quantum
estimation theory \cite{helstrom76,jones91} as was done
in Ref.~\cite{rehacek99b,hradil96,zawisky98}.
Let us consider the 8-port homodyne detection scheme
\cite{noh92,noh91} with four output channels
numbered by indices 3,4,5,6, where the actual values of
intensities are registered in each run. Assume that
these values fluctuate in accordance with some statistics.
The mean intensities are  modulated by a phase parameter
$\bar{\theta}$
\begin{eqnarray} \label{stredni}
\bar n_{3,4} &=& \frac{N}{2} (1\pm V \cos\bar{\theta}), \nonumber\\
\bar n_{5,6} &=& \frac{N}{2} (1\pm V \sin \bar{\theta}),
\end{eqnarray}
where $N$ is the total intensity and $V$ is the visibility of
interference fringes.
The true phase shift inside the interferometer $\bar{\theta}$,
which is a nonfluctuating parameter controlled by the
experimentalist, should
carefully be distinguished from the estimated phase shift,
which is a random quantity. Hereafter, the latter is denoted
by $\theta$ .
This device, operating with Gaussian signals, represents
nothing but a classical wave
picture of the original 8-port homodyne detection scheme.
Equivalently, it also corresponds to
a Mach-Zehnder interferometer, when the measurement
is performed with zero  and $\pi/2$
auxiliary phase shifters.
In this case, data is not obtained simultaneously,
but it is collected during repeated experiments.
Provided that a particular combination of outputs
$\{n_3,n_4,n_5,n_6\}$
has been registered, the phase shift can be inferred.
The  point estimators  of phase corresponding to
the maximum--likelihood  (ML)
estimation  will be used here \cite{lane93,hradil97}.
In accordance with the ML approach \cite{kendall61},
the sought-after  phase
shift is given by the  value, which maximizes the likelihood
function.
Provided that the noise is Gaussian, the likelihood function
corresponding to the detection of given data reads
%%%%%%%%%%%%%%%%%%%%%%%%%%%%%%%%%%%%%%%%%%%%%%%%%%%%%%%%%%%%%%%%%%%%%%
\begin{eqnarray}
\label{LGauss1}
{\cal  L}\propto
           \exp\left\{-\frac{1}{2\sigma^2} \sum_{i=3}^{6}
           [n_i - \bar n_i ]^2\right\} .
\end{eqnarray}
%%%%%%%%%%%%%%%%%%%%%%%%%%%%%%%%%%%%%%%%%%%%%%%%%%%%%%%%%%%%%%%%%%%%%%
Here the variance $\sigma^2$  represents the phase insensitive
noise of each channel.
A notation analogous to the definition of phase by Noh, Fouger\`es
and Mandel \cite{noh93} may be introduced
%%%%%%%%%%%%%%%%%%%%%%%%%%%%%%%%%%%%%%%%%%%%%%%%%%%%%%%%%%%%%%%%%
\begin{eqnarray} \label{nfm_estim}
 e^{i \theta_{N\!F\!M}}= \frac{n_3-n_4 + i (n_5-n_6)}
{\sqrt{(n_3-n_4)^2 + (n_5-n_6)^2}},
\\
V' =  \sqrt{(n_3-n_4)^2 + (n_5-n_6)^2}.
\end{eqnarray}
%%%%%%%%%%%%%%%%%%%%%%%%%%%%%%%%%%%%%%%%%%%%%%%%%%%%%%%%%%%%%%%%%
The likelihood function (\ref{LGauss1})
may be  maximized by the choice of
parameters for phase shift and visibility,
respectively \cite{rehacek99b}
%%%%%%%%%%%%%%%%%%%%%%%%%%%%%%%%%%%%%%%%%%%%%%%%%%%%%%%%%%%%%%%%%
\begin{eqnarray}  \label{wave}
\theta = \theta_{N\!F\!M},
\label{NFM1} \\
V = \min \biggl( \frac{2 V'}{\sum_{i=3}^6 n_i}, 1 \biggr).
\label{NFM2}
\end{eqnarray}
%%%%%%%%%%%%%%%%%%%%%%%%%%%%%%%%%%%%%%%%%%%%%%%%%%%%%%%%%%%%%%%%%
Hence  the operational phase concept  of Noh, Fouger\`es and Mandel
coincides with the ML estimation  for waves represented by
continuous  Gaussian signal with phase independent and symmetrical
noises. These rather strict assumptions are incompatible with
the nature of signals encountered in experiments; such restrictions
would be, however, natural in the classical theory.

The optimum prediction is different for  Poissonian statistics. ML
estimation based on the Poissonian likelihood function
%%%%%%%%%%%%%%%%%%%%%%%%%%%%%%%%%%%%%%%%%%%%%%%%%%%%%%%%%%%%%%%%
\begin{eqnarray}
\label{lik}
&&{\cal L}\propto
\prod_{i=3}^6\bar{n}_i^{n_i}
\end{eqnarray}
%%%%%%%%%%%%%%%%%%%%%%%%%%%%%%%%%%%%%%%%%%%%%%%%%%%%%%%%%%%%%%%%
gives optimum values for phase shift and
visibility \cite{rehacek99b}
%%%%%%%%%%%%%%%%%%%%%%%%%%%%%%%%%%%%%%%%%%%%%%%%%%%%%%%%%%%%%%%%
\begin{eqnarray}
  \label{phase_p}
e^{i\theta}=\frac{1}{V}
 \left[
   \frac{n_4-n_3}{n_4+n_3}+i\frac{n_6-n_5}{n_6+n_5}
 \right],
\label{max1} \\ \label{vis_p}
V~=\sqrt{{\left( \frac{n_4-n_3}{n_4+n_3}\right)}^2+
        {\left( \frac{n_6-n_5}{n_6+n_5}\right)}^2 },
\label{max2}
\end{eqnarray}
%%%%%%%%%%%%%%%%%%%%%%%%%%%%%%%%%%%%%%%%%%%%%%%%%%%%%%%%%%%%%%%
provided the estimated visibility (\ref{vis_p}) is smaller
than unity.
In the opposite case it is necessary to maximize the
likelihood function (\ref{lik})
on the boundary ($V=1$) of the physically allowed region
of the parameter space numerically.
Relations (\ref{phase_p}-\ref{vis_p}) provide a  correction  of the
Gaussian theory with
respect to the discrete signals.
Besides the phase shift, visibility  of
interference fringes and the total input energy
can be evaluated simultaneously.

The apparent difference between relations (\ref{NFM1}--\ref{NFM2})
and (\ref{max1}--\ref{max2})  represents the theoretical
background  of the presented treatment.
Obviously, both predictions will coincide provided that there is
almost no information  available in the low field limit $N
\rightarrow 0.$
Similarly in the strong field limit $N \rightarrow \infty ,$  the
phase of the light is well defined and both inferred values of the
phase approach the same value. Possible deviations  may appear
in the intermediate regime $N \approx 1$.
The test of the difference  between  $(\ref{NFM1}) $ and
$(\ref{max1})$ is proposed  as controlled   phase measurement.
The phase difference was adjusted to a certain value and
estimated independently using  both methods
 $(\ref{NFM1}) $
and $(\ref{max1})$ in repeated experiments.

To compare two or more phase estimators, some measure of the
estimation error is needed. Dispersion defined as
%%%%%%%%%%%%%%%%%%%%%%%%%%%%%%%%%%%%%%%%%%%%%%%%%%%%%%%%%%%%%%%%%%%%%
\begin{equation} \label{disp}
\sigma^2=1-\left|\langle \mbox{e}^{i\theta}\rangle\right|^2
\end{equation}
%%%%%%%%%%%%%%%%%%%%%%%%%%%%%%%%%%%%%%%%%%%%%%%%%%%%%%%%%%%%%%%%%%%%%
can well do the job. Here the average is taken over posterior phase
distribution of the corresponding phase estimator.
The dispersion (\ref{disp}) is a compact space analogy of
the averaged quadratic cost function (variance),
frequently used in the estimation theory
\cite{helstrom76}.

The evaluation of the average quadratic cost (\ref{disp}) is not
the only way to compare efficiencies of different estimation
procedures. Another possibility is to use the rectangular cost function
%%%%%%%%%%%%%%%%%%%%%%%%%%%%%%%%%%%%%%%%%%%%%%%%%%%%%%%%%%%%%%%%%%%
\begin{equation}\label{rect_cost}
C(\theta-\bar{\theta})=
\left\{
\begin{array}{rcccc}
-1&\,\,&|\theta-\bar{\theta}|&\leq&\Delta\theta\\
0&\,\,&|\theta-\bar{\theta}|&>&\Delta\theta\\
\end{array}\right.
.
\end{equation}
%%%%%%%%%%%%%%%%%%%%%%%%%%%%%%%%%%%%%%%%%%%%%%%%%%%%%%%%%%%%%%%%%%%
This choice of the cost function corresponds to the following evaluation
the experimental data. Each sample of data consisting of numbers
$n_3,n_4,n_5,n_6$ of
counted photons is processed using NFM formula (\ref{nfm_estim})
issuing phase prediction $\theta_{N\!F\!M}$. The relative
frequency $f_g(\Delta\theta)$,
which is proportional to the average cost of
the Gaussian estimator
$\langle C(\theta-\bar{\theta})\rangle$,
characterizes  how many times  the
estimated phase $\theta_{N\!F\!M}$ falls within the chosen phase
window $\Delta\theta$ (confidence interval) spanning around the true phase
shift. The same procedure is repeated for phase predictions based
on the Poissonian phase estimator (\ref{phase_p})
yielding the relative frequency of
``hits'' $f_p(\Delta\theta)$. The quantity
%%%%%%%%%%%%%%%%%%%%%%%%%%%%%%%%%%%%%%%%%%%%%%%%%%%%%%%%%%%%%%%%%%%%
\begin{eqnarray} \label{diff}
\Delta E&=&f_P(\Delta\theta)-f_G(\Delta\theta)
\end{eqnarray}
%%%%%%%%%%%%%%%%%%%%%%%%%%%%%%%%%%%%%%%%%%%%%%%%%%%%%%%%%%%%%%%%%%%%
represents the difference in efficiency of the ML and NFM phase
estimations for the given  phase window  $\Delta\theta$ and given
input energy $N$. If this quantity is significantly
positive, the ML estimation
is better than its NFM counterpart.
On the other hand, if $\Delta E$ is close to zero,
both data evaluation procedures are statistically
equivalent and no discrimination is possible.

\section{Experimental setup} \label{sec_setup}

\begin{figure}
  \vspace{-0.6cm}
  \centerline{\epsfxsize=9cm \epsfbox{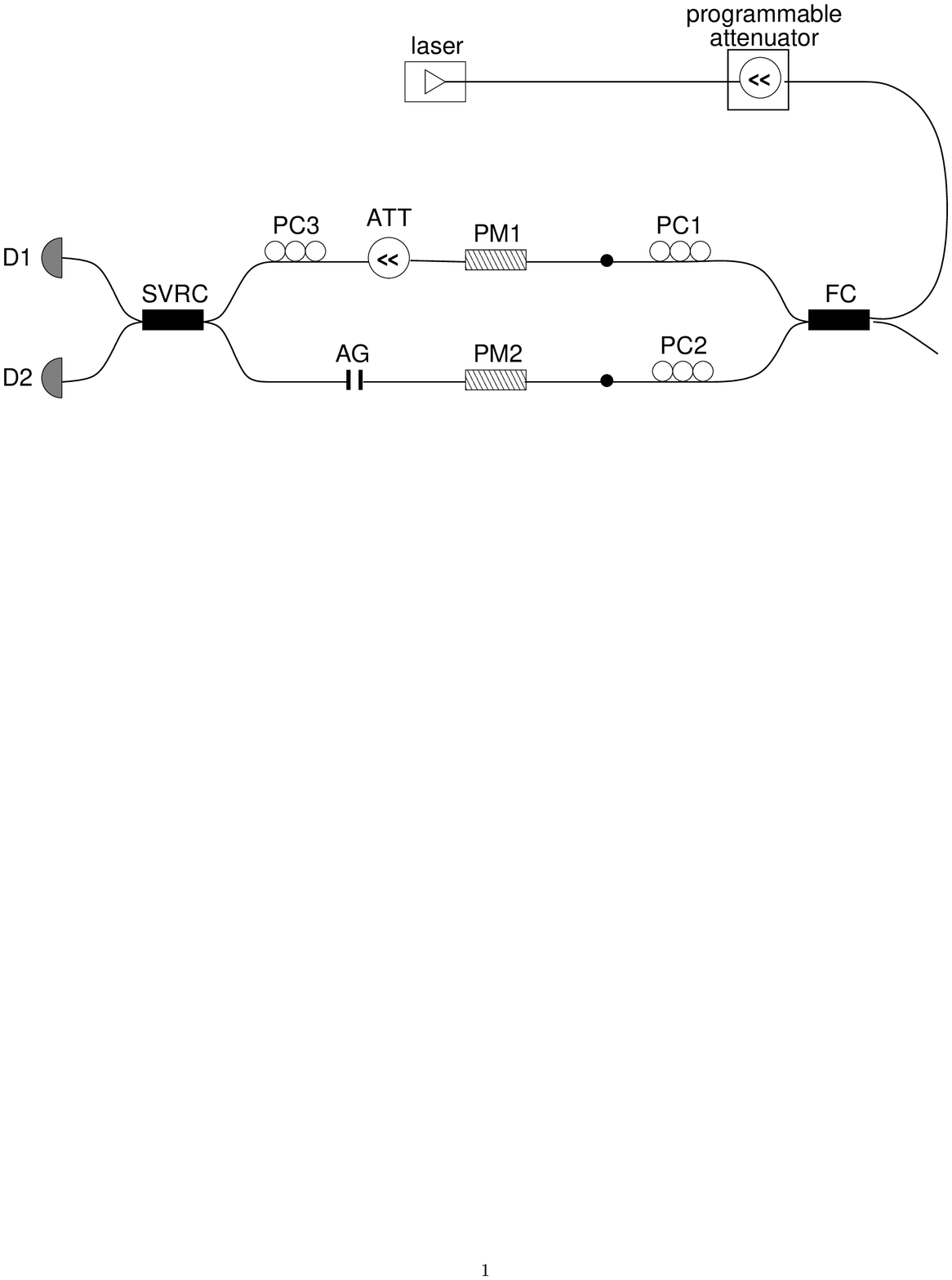}}
  \vspace{-7.5cm}
\caption{Scheme of the laboratory setup. FC - input fiber coupler,
PCx - polarization controllers, PMx - phase modulators,
ATT~-~attenuator, SVRC - output variable ratio coupler,
Dx - detectors.} \label{Fig_setup}
\end{figure}

The laboratory setup (see Fig.~\ref{Fig_setup}) is based on a
single--mode-fiber Mach-Zehnder interferometer
carefully balanced and adjusted for maximum visibility.
A semiconductor laser source (SHARP LT015) produces 4--ns--long pulses
with a repetition rate of 130 kHz. Initial pulse intensity is about
\( 10^{7} \) photons per pulse. This is decreased
by 11 dB due to losses in the fibers and other components
of the setup, and precisely adjusted by artificial attenuation in
the programmable attenuator (JDS Fitel HA9) to reach the required
level at the detectors. Input coupler FC (SIFAM) divides the pulses
between the arms of the interferometer (each 4 m long). Both arms
contain planar phase modulators PM1,2 (UTP). Only PM1 has been used
for phase settings, the other is included just for symmetry reasons.
Both modulators also work as linear polarizers (extinction ratio
\( 1:10^{6} \)) to improve the degree of polarization. Input polarization
to the modulators is set by polarization controllers PC1 and PC2.
Attenuator ATT in the upper arm of the interferometer
helps balance the losses in both arms of the interferometer to
reach maximum visibility. The length of the arms is balanced by a
variable air gap (AG). Polarization controller PC3 is used to match
the polarization in the arms at the output variable
ratio coupler SVRC (SIFAM). The result of interference is detected
using silicon photon-counting detectors (EG\&G SPCM-AQ) with less than
70 dark counts per second and quantum efficiency of 55\%. The signals from
the detectors are processed using detection electronics based on
time-to-amplitude converters and single-channel analyzers
(EG\&G Ortec) and recorded
by a computer, which also controls the driving voltage of the phase
modulator, programmable attenuator setting and laser operation as well.
In this setup we have reached interference
visibility of up to 99.8\%.

The whole interferometer is placed in a polystyrene box to minimize
thermal drift of the fringes. After initial warm-up, the phase stability
of the device is better than \( \pi /3000 \) per second. During the
measurement, active stabilization of the interference pattern is performed
each 5-10 s.

\section{Measured data evaluation} \label{sec_evaluation}

Unfortunately, commercially available photodetectors
for measurement of weak quantum signals fail to
discriminate the number of detected photons. Only the presence
or absence of the signal can usually be detected.
The impossibility to count photons is circumvented as follows.
According to the well--known polynomial theorem, the sum of two or more
Poissonian signals is a Poissonian signal again, the mean simply being
replaced by the sum of the means of its constituents.
\begin{figure}[t]
  \vspace{-1cm}
  \centerline{\epsfxsize=10cm \epsfbox{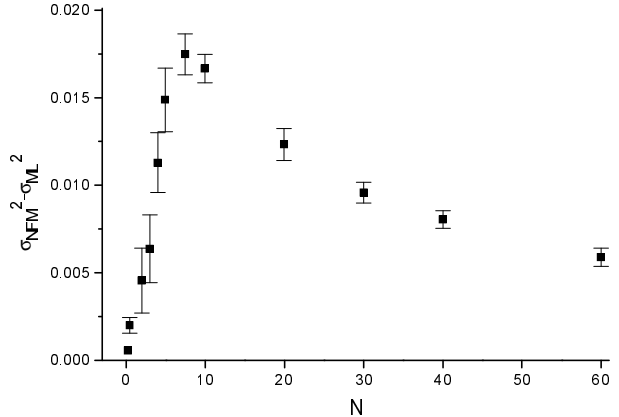}}
  \vspace{-7.8cm}
  \caption{The experimentally observed difference between dispersions
  of the
NFM and ML estimators as a function of the
input mean number of photons $N$ for
fixed true phase $\bar{\theta}=\pi/3$.
Error bars corresponding to
$68$\% confidence intervals are shown.} \label{Fig_disp}
  \vspace{0cm}
\end{figure}
It is therefore possible to
carry out measurements with very weak signals of intensity, say,
$0.01-0.001$ photons per pulse so that probability of two photons
being in the same pulse (double--detection)
is very small, and then collect an appropriate number of individual
yes--no detections to obtain desired ``input'' intensity $N$.
For example, an experimental run with input pulse mean intensity
$N=10$ can be simulated by a sequence of $10000$ measurement
with mean input intensity $N_p=0.001$ photons per pulse.
The probability of double--detection in a single run is
$p<10^{-6}$ for a Poissonian light source.
Hence the probability of single double--detection
during the whole sequence of measurements
is less than $1\%$ and the probability
of triple--detection or several double--detections is
entirely negligible.
This procedure enables us to effectively
simulate the results of experiments with
intense pulses $N\gg1$ and ideal photodetectors. Whenever in the
text an experimental sample is mentioned, it should be clear that
we actually refer to a sum of many experimental samples measured
with intensities well below a single photon per pulse.

The difference of dispersions (\ref{disp}) of the Gaussian
and Poissonian phase estimators found in our experiment
is shown in Fig.~\ref{Fig_disp}
for a fixed true phase $\bar{\theta}=\pi/3$.
The number of
experimental samples used for calculation of the dispersions varies
from $1000$ samples for input intensity $N=60$ to more than
$100,000$ samples with $N=0.1$. The error bars arising from
a limited number of samples are the result of numerical simulation.
Fig.~\ref{Fig_disp} agrees well with qualitative reasoning of the
previous section. The most distinct difference between
the dispersions of the ML and NFM estimators is seen for the
input mean number of photons $N\approx 7.5$. Thus, it can be said
that as long as interference and phase measurements are concerned,
discrete signals with Poissonian statistics are
distinguishable from the classical wave for only
a relatively small range of input energies.

\begin{figure}
  \vspace{-0.7cm}
  \centerline{\epsfxsize=10.5cm \epsfbox{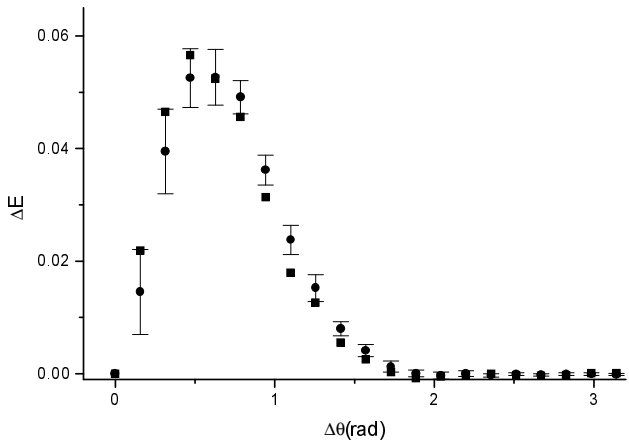}}
  \vspace{-8.3cm}
\caption{
Experimentally obtained $\Delta E$
(squares) compared to
theoretical values (circles).
Error bars corresponding to $7500$ measured samples are shown.}
\label{Fig_delta_E}
  \vspace{0cm}
\end{figure}
The difference in efficiency of the ML and NFM phase estimation
(\ref{diff}) calculated from experimental data is shown in
Fig.~\ref{Fig_delta_E}.
The difference $\Delta E$ was calculated using $7500$
experimental samples measured in experiment with
$N=10$ photons and visibility of $99.6\%$. The chosen input
energy roughly corresponds to the maximum seen in
Fig.~\ref{Fig_disp}.
Since the  experimental data are limited to a finite
number of samples due to experimental conditions and
available time, the estimated $\Delta E$
would be slightly different in repeated experiments.
Statistical significance of the
experimental results is demonstrated using computer
simulation again. Standard deviation corresponding to
$7500$ measured samples is shown in Fig.~\ref{Fig_delta_E}
as error bars for each phase window.

A significant difference between the effectiveness of  classical
and optimal  treatments is apparent in the  Fig.~\ref{Fig_delta_E}.
The optimal
treatment provides an improvement in estimation procedure,
and the  difference is beyond the statistical error by more
than $10$ standard deviations in the optimum case. High stability and
visibility of interference fringes in the optical
interferometer along with a high repetition rate of the pulsed laser
make the improvement of the NFM phase prediction
more evident than in a similar comparison performed with
the neutron interferometer setup \cite{rehacek99b}.
Notice the dependence of the precision gain $\Delta E$ on the
width of the chosen phase window.
\begin{figure}
  \vspace{-1cm}
  \centerline{\epsfxsize=10cm \epsfbox{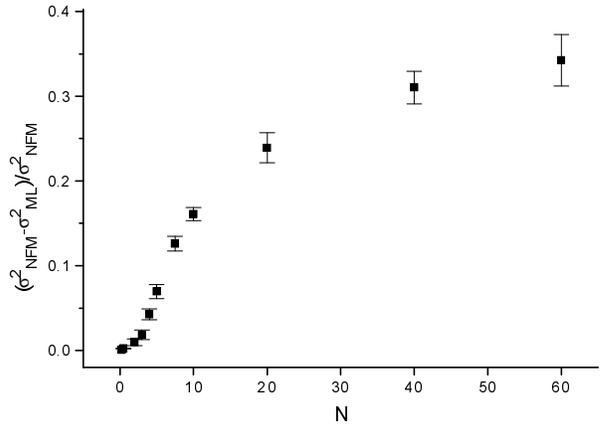}}
  \vspace{-7.8cm}
\caption{Observed relative difference between dispersions of the
NFM and ML estimators. All the parameters are the same as
in Fig.~\protect{\ref{Fig_disp}}.} \label{Fig_disp_rel}
  \vspace{0cm}
\end{figure}
Obviously, no better performance of the ML method
can be expected for large values of
the phase window $\Delta\theta$; any sensible statistical method
would yield quite reasonable results. Likewise, no real
improvement over the Gaussian estimate can be expected when
$\Delta\theta$ is close to zero, because too few data would then
fall within the window. The largest difference is
about $6\%$  in the window of the width of about $0.5$ rad.

\section{Asymptotic behavior} \label{sec_asymptotic}

As intensity of the input light increases, both estimations of
phase shift yield more sharp and precise results.
The error of the
best--known proposed phase measurements scales as
$N^{-1}$ for large input
intensities \cite{holland93,yurke86}.
Our photodetection scheme cannot
compete with such measurements. On the other hand
these methods necessitate the use of exotic states such as two--mode
Fock states, etc., which are still impossible to prepare in
contemporary laboratories.
Since the experimental equipment is always limited,
the only way to improve precision of phase measurements
lies in careful evaluation of
the measured phase sensitive data and distillation of all available
phase information.
Therefore, it is worthwhile to compare the performance of the NFM and
ML estimations in the limit of high intensities.

To get some
qualitative feeling of how both estimators approach
the above mentioned limit we redrew Fig.~\ref{Fig_disp}
with differently scaled vertical axes, see Fig.~\ref{Fig_disp_rel}.
It can be seen that the relative difference of both dispersions
monotonically increases with $N$ and finally approaches some constant
value different from zero. This means that during the transition from
$N\approx 10$ to higher values of $N$, both estimators first scale
with slightly different powers of $N$, and for high intensities
both powers reach the same value and the ratio of the Gaussian and
Poissonian dispersions approach some constant.

It is easy to calculate the behavior of the dispersion of the Gaussian
phase estimator for $N\rightarrow\infty$. The dispersion becomes
%%%%%%%%%%%%%%%%%%%%%%%%%%%%%%%%%%%%%%%%%%%%%%%%%%%%%%%%%%%%%%%%%
\begin{equation}\label{asym_g}
\sigma^2_{G}\approx\frac{1}{V^2}N^{-1}+O
\left(\frac{1}{N^2}\right).
\end{equation}
%%%%%%%%%%%%%%%%%%%%%%%%%%%%%%%%%%%%%%%%%%%%%%%%%%%%%%%%%%%%%%%%%
Readers interested in details are referred to Appendix~\ref{ap_1}.
As can be expected, the error of NFM phase measurement
is proportional to $N^{-1/2}$. This precision represents the
so--called standard quantum limit. Unfortunately, it is
impossible to derive a simple expression similar to Eq.~(\ref{asym_g})
for the Poissonian estimator. However, such a
formula is easily obtained provided the physical constraint
$V\leq1$ is released.
Thus throwing the estimated value of
visibility away and interpreting Eq.~(\ref{phase_p}) as an
estimator of the unknown phase shift valid for each sample
$\{n_3,n_4,n_5,n_6\}$, the asymptotic dispersion of such
an unconstrained estimation reads
%%%%%%%%%%%%%%%%%%%%%%%%%%%%%%%%%%%%%%%%%%%%%%%%%%%%%%%%%%%%%%%%%
\begin{equation}\label{asym_p}
\sigma^2_{P}\approx\frac{1}{V^2}\left(1-\frac{V^2}{2}
\sin^2 2\bar{\theta}\right)N^{-1}+O\left(\frac{1}{N^2}\right).
\end{equation}
%%%%%%%%%%%%%%%%%%%%%%%%%%%%%%%%%%%%%%%%%%%%%%%%%%%%%%%%%%%%%%%%%
It is obvious that omitting useful information gained
from the data makes the unconstrained estimation somewhat
less efficient than the original constrained one.

We can see from Eqs. (\ref{asym_g}) and (\ref{asym_p}) that
in the limit of low visibility both the NFM and ML phase
predictions are equivalent. This result is in agreement with
properties of the well--known discrete Fourier transform
(DFT) phase estimator \cite{goodman73} in the same
regime\footnote{We note in passing that DFT can in fact
be regarded as
a generalization of NFM phase concept to a greater number of
auxiliary phase shifts \cite{rehacek99b}.}.
On the other hand, ML estimation always
gives better results than NFM theory, provided the visibility
is high.
For some values of the true phase
shift, reduction in the dispersion down to
$50\%$ is possible.

The ultimate limit to the resolution of the particular estimator is
set by the well known Cram\'{e}r--Rao inequality. Provided the
visibility and input intensity are under control in the experiment,
the phase shift $\bar{\theta}$ remains
the only parameter to be estimated.
For such a single--parameter problem, the
Cram\'{e}r--Rao lower bound (CRLB) on the
dispersion\footnote{Actually, the CRLB holds for variance
rather than for dispersion. Since all the relevant
phase uncertainties are small in the limit of high
input energy, both quantities coincide in this case.}
of any phase estimator is given as follows
%%%%%%%%%%%%%%%%%%%%%%%%%%%%%%%%%%%%%%%%%%%%%%%%%%%%%%%%%%%%%%%
\begin{equation}\label{crlb}
\sigma^2_{CRLB}=\left(
E\left\{\left[(\partial/\partial\bar{\theta})\ln
p({n_3,n_4,n_5,n_6}|\bar{\theta})\right]^2\right\}\right)^{-1}.
\end{equation}
%%%%%%%%%%%%%%%%%%%%%%%%%%%%%%%%%%%%%%%%%%%%%%%%%%%%%%%%%%%%%%%
Here the symbol $E$ denotes averaging over observed data.
\begin{figure}[t]
  \vspace{-1cm}
  \centerline{\epsfxsize=10cm \epsfbox{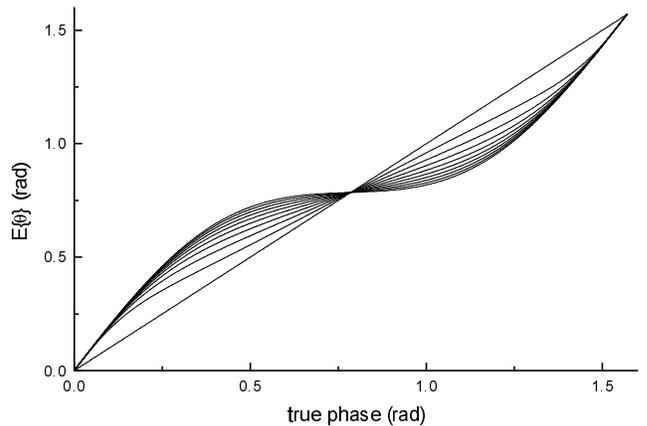}}
  \vspace{-7.8cm}
\caption{Bias of the single--parameter phase estimator for expected
visibility $V$=$1$ as a function of the true phase shift
$\protect{\bar{\theta}}$.
Actual visibility of interference fringes varies from $V$=$1$
(straight line) to $V$=$0.1$ (most bent line) in steps
$\Delta V$=$0.1$.}\label{Fig_bias}
  \vspace{2mm}
\end{figure}
Upon substitution of the joint Poissonian distribution of the
sample $\{n_3,n_4,n_5,n_6\}$ to Eq. (\ref{crlb}) and making similar
approximations to those used in derivation of Eq.~(\ref{asym_g}),
we end up with
%%%%%%%%%%%%%%%%%%%%%%%%%%%%%%%%%%%%%%%%%%%%%%%%%%%%%%%%%%%%%%%
\begin{equation}\label{asympt_crlb}
\sigma^2_{CRLB}=
\frac{V^2-1-\frac{1}{4}V^4\sin^22\bar{\theta}}
{V^2-1-\frac{1}{2}V^2\sin^22\bar{\theta}}V^{-2}N^{-1}.
\end{equation}
%%%%%%%%%%%%%%%%%%%%%%%%%%%%%%%%%%%%%%%%%%%%%%%%%%%%%%%%%%%%%%%
Two interesting observations follow from (\ref{asympt_crlb}).
First, notice that for low visibility, the dispersion of the NFM
phase prediction (\ref{asym_g}) attains the CRLB.
This means the NFM estimation is best possible in this limit.
Second, for perfect experimental setup ($V$=$1$), the CRLB is
simply $\sigma^2_{CRLB}=1/2N$.
%%%%%%%%%%%%%%%%%%%%%%%%%%%%%%%%%%%%%%%%%%%%%%%%%%%%%%%%%%%%%%%%%
\begin{table}
\begin{tabular}{lrr}
\hline
Estimator & $\sigma^2$ &
$\bar{C}\equiv\int\sigma^2 d\bar{\theta}$ \\
\hline
NFM & $1/N$& $2\pi/N$ \\
unconstr. ML & $\left(1+\cos^2 2\bar{\theta}\right)\!\!/2N$ &
$\frac{3}{2}\pi/N$\\constr. ML &
$\approx\left(1+0.5\cos^2 2\bar{\theta}\right)\!\!/2N$
& $\approx \frac{5}{4}\pi/N$\\
CRLB & $1/2N$& $\pi/N$ \\
\hline
\end{tabular}
\caption{Asymptotic dispersion and overall quadratic cost of various
phase estimators. For comparison, CRLB is shown. Note that
phase prediction of the ML estimation with physical constraint on the
inferred value of visibility is superior to the prediction without the
constraint.} \label{summary}
%\vspace*{-3mm}
\end{table}
%%%%%%%%%%%%%%%%%%%%%%%%%%%%%%%%%%%%%%%%%%%%%%%%%%%%%%%%%%%%%%%%%
Thus the resolution of the
unconstrained ML estimation can
still be improved a bit.
In deriving Eq. (\ref{asympt_crlb}) we supposed the
value of visibility and input intensity of the laser beam are known.
Such knowledge represents some additional information about the
experimental setup. Let us assume that the input intensity and
visibility are really under control and, in addition, the visibility
is equal to unity. In this case the Poissonian likelihood function
(\ref{lik}) depends only on the value of the phase shift
$\bar{\theta}$. Now the single--parameter ML estimation of
the phase shift $\bar{\theta}$
consists of maximizing the likelihood function
${\cal L}(\bar{\theta},V$=$1)$
with respect to the
single parameter $\bar{\theta}$.
This procedure is just what we have done
in the case of many--parameter ML estimation
(\ref{phase_p}-\ref{vis_p}), when the experimental sample yielded
an unphysical value of visibility $V>1$. The only difference is
that in the case of single--parameter ML estimation
we maximize the likelihood function on the boundary
for any detected sample $\{n_3,n_4,n_5,n_6\}$. We may ask whether
the single--parameter phase estimator achieves the best phase
resolution $\sigma^2_{CRLB}$=$1/2N$.
An explicit calculation (for details see Appendix~\ref{ap_2})
shows this is really the case. Although it may seem that the
single--parameter ML estimation thus gives best results, some caution
is necessary when visibility (or another parameter) is not known precisely
or fluctuates. For example, estimation on the boundary $V=1$ leads
to a strongly biased phase prediction provided the actual visibility
differs from unity, as is demonstrated in Fig.~\ref{Fig_bias}.
In this particular case, the bias caused by dismissing
the possibility $V<1$ is independent of the input light
intensity $N$. For larger intensities it dominates the
uncertainty of estimated phase and the single--parameter ML
estimation may be outperformed by the Gaussian (NFM) one.
Therefore one should always
estimate all parameters, which are not
under experimentalist's control together with the
parameter of interest regardless of smaller
theoretical effectiveness
of such a complex estimation procedure.

Now let us return to the many--parameter constrained ML estimation.
In this case a particular detection \{$n_3$, $n_4$, $n_5$, $n_6$\} is
processed either by Eq.~(\ref{phase_p}) (when applied to all samples,
$\sigma^2$ is given by Eq.~(\ref{asym_p})) or via maximization of
likelihood function (\ref{lik}) on the boundary (when applied to all
samples, $\sigma^2$=$1/2N$). Although we do not switch between these
methods at random, because the choice depends on the particular
sample, the mean of both dispersions gives us a rough estimate
of the performance of the constrained ML phase prediction.
A more precise value can always be obtained with the help of computer
simulation.
The performances of various phase estimators are
summarized in Tab.~\ref{summary} for a perfect
experimental setup $V$=$1$.

\begin{figure}
  \vspace*{-0.6cm}
  \centerline{\hspace{4mm}\epsfxsize=10.9cm \epsfbox{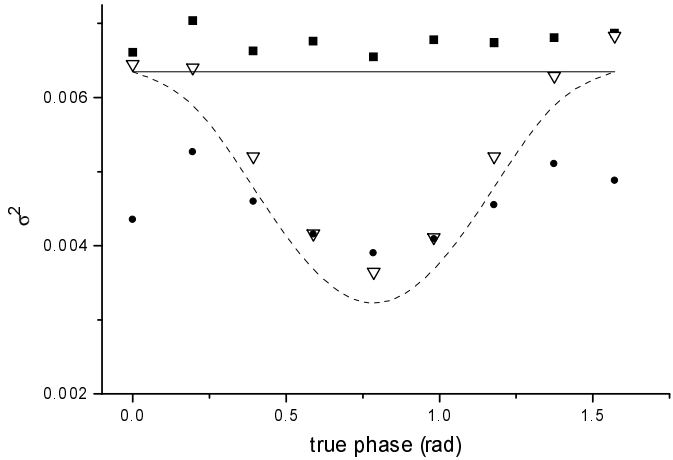}}
  \vspace*{-8.5cm}
\caption{Asymptotic dispersion of the NFM estimator;
theory (solid line) and experimentally obtained values (squares).
Asymptotic dispersion of the unconstrained ML estimator;
theory (dashed line) and experimentally obtained values (triangles).
Experimentally obtained dispersion of the constrained ML estimation
(circles). The corresponding input mean number of photons and
the estimated visibility are $N=160$ and
$V=99.2\%$, respectively.}
\label{Fig_disp_asymptot}
  \vspace{0mm}
\end{figure}

An experimental comparison of the three phase estimations in the
asymptotic regime is shown in Fig.~\ref{Fig_disp_asymptot}.
For comparison, the theoretical values of dispersions given by
Eqs.~(\ref{asym_g}) and (\ref{asym_p}) are also shown.
The dispersions were determined using $10,000$ measured samples with
$N=160$ for each value of the true phase shift
$\bar{\theta}=k\pi/16~$rad, $k=0,1,..,8$.
More than $10^9$ weak laser pulses were sent through
the interferometer to obtain the figure.
Several important conclusions can be drawn from
Fig.~\ref{Fig_disp_asymptot}.
(i) We can see that the uncertainty of the constrained
ML estimation is definitely below the uncertainty of the
unconstrained estimation in agreement with our arguments presented
in this section. It means that insisting on the physical
constraints\footnote{here non--negative definiteness
of the intensity}
of allowed results of estimation or reconstruction procedure
is important not only for interpretation
reasons, but it also makes the estimation more efficient.
(ii) The observed values of dispersion exhibit a systematic error.
The additional noise above the theoretical uncertainty is caused
by inherent phase fluctuations in the experimental setup and
their magnitude can be estimated from Fig.~\ref{Fig_disp_asymptot}
as $0.020\pm 0.003$ rad. This value is in an excellent agreement with
the value $0.019$ rad obtained by an independent method.
Hence our statistically motivated evaluation of experimental data
can be used for inferring the amount of fluctuations, and therefore
it provides an independent and nontrivial way for calibrating
an interferometer. Moreover, a slightly different sensitivity of
different phase estimators to various parameters of the setup
makes it possible, at least in principle, to distinguish between
different sources of noise.
This is another interesting feature
of the method we propose.
(iii)
In Section~\ref{sec_evaluation} we could see that the
most distinct difference between semiclassical and
fully quantum phase concepts occurs in the regime, where the
intrinsic phase uncertainty of light is much larger than phase
fluctuations caused by any reasonable imperfections of the
experimental setup. Therefore, though clearly visible in
Fig.~\ref{Fig_disp_asymptot},
the ``external'' phase fluctuations may be completely neglected
in Fig.~\ref{Fig_disp}. However, with increasing intensity the
(unavoidable) fluctuations become comparable with the intrinsic phase
uncertainty, and for even larger $N$ the accuracy of any phase
measurement is governed by the external influences rather than by
the theoretical limit of the corresponding phase estimation.
The statistics of light are then no longer reflected in its phase
properties, and different quantum phase concepts become
indistinguishable.
Not only does this provide another evidence
for the fact that the NFM phase concept differs from its ML counterpart
only for a narrow range of energies, as we already stated in
section~\ref{sec_evaluation},
but it also shows how the operationally defined quantum phase
approaches its classical limit.

\section{Conclusion}

Theoretical and experimental justification of operational quantum
phase concepts is addressed in this paper. Statistically motivated
evaluation of the  interferometric setup has been presented. The
choice of the optimum phase estimator strongly depends on the
experimentalist's knowledge about the interferometric setup and on
the nature of the signal being detected. Two important cases
--  NFM and  ML estimators, resulting from the
classical and quantum description of the experiment, respectively,
have been compared. Differences between both  treatments have
 been measured experimentally and have shown to be statistically
significant in the limited range of input energies. In particular,
 no difference between the NFM and ML phase predictions have been
 observed in the regime of very small number of particles, which is
usually considered as the domain of quantum physics.
 This detailed  analysis makes it possible quantify the amount of
 the noise associated with the  phase. The lack of knowledge about
the parameters of interferometric setup has also been considered.
In the asymptotic limit of large input energy the intrinsically
biased ML estimation procedures yield sensible results only
provided that all the uncertain parameters of the setup are
estimated together with the unknown phase shift. This can be
interpreted  in the framework  of more complex estimation
procedures -- the so called quantum  state reconstructions.

\vspace*{10pt}
\noindent
{\large \bf Acknowledgments}

We acknowledge support by the TMR Network ERB FMRXCT 96-0057 "Perfect
Crystal Neutron Optics" of the European Union,
by grant No VS96028 and by research project
CEZ:J14/98 ``Wave and particle optics'' of
the Czech Ministry of Education.

\appendix

\section{Asymptotic \, dispersion\,
of \, the Gaussian (NFM) estimator}
\label{ap_1}

To calculate the dispersion of the phase estimator, we need to evaluate
the expectation of the sine and cosine functions of the inferred phase,
e.g.
%%%%%%%%%%%%%%%%%%%%%%%%%%%%%%%%%%%%%%%%%%%%%%%%%%%%%%%%%%%%%%%%%
\begin{equation} \label{expect}
\langle \cos\theta\rangle = \hspace{-2mm}\sum_{n_3,..,n_6}
\hspace{-2mm}
\cos\left[\theta(n_3,n_4,n_5,n_6)\right] \prod_{i=3}^6
P(n_i).
\end{equation}
%%%%%%%%%%%%%%%%%%%%%%%%%%%%%%%%%%%%%%%%%%%%%%%%%%%%%%%%%%%%%%%%%
Here the inferred phase shift $\theta$ conditioned by the detection
($n_3,n_4,n_5,n_6$) is given by Eqs. (\ref{NFM1}) and
(\ref{phase_p}) for the NFM and ML estimations, respectively. Now,
suppose the interferometer is fed by a strong pulse with $N\gg 1$.
Provided the true phase shift $\bar{\theta}\neq k\pi/4, \,
k\in\cal{N}$,
we also have $\bar{n}_i\gg 1, \, i=3,4,5,6$, and the Poissonian
photocount distribution $P(n_i)$ can be approximated by Gaussian
with the same variance, near its peak:
%%%%%%%%%%%%%%%%%%%%%%%%%%%%%%%%%%%%%%%%%%%%%%%%%%%%%%%%%%%%%%%%%
\begin{eqnarray}\label{approx}
P(n_i)=\frac{\bar{n}_i^{n_i}}{n_i!}e^{-\bar{n}_i}\approx
\frac{e^{-(n_i-\bar{n}_i)^2/2\bar{n}_i}}{\sqrt{2\pi\bar{n}_i}},\\
\bar{n}_i\gg 1, \,\, n_i-\bar{n}_i \ll \bar{n}_i.
\end{eqnarray}
%%%%%%%%%%%%%%%%%%%%%%%%%%%%%%%%%%%%%%%%%%%%%%%%%%%%%%%%%%%%%%%%%
Notice that although the distribution now becomes symmetric (i.e.,
the estimation is unbiased for large $N$), the noise remains phase
sensitive even for high input energy in contrast to the assumption
hidden in the NFM theory (\ref{LGauss1}). Using the Gaussian phase
formula (\ref{nfm_estim}) in (\ref{expect}), we obtain
the expectation value of the cosine phase function
in the following form
%%%%%%%%%%%%%%%%%%%%%%%%%%%%%%%%%%%%%%%%%%%%%%%%%%%%%%%%%%%%%%%%%
\begin{equation}\label{gauss_int}
\langle\cos\theta\rangle=\int\frac{n_{34}}
{\sqrt{n^2_{34}+n^2_{56}}} P(n_{34})P(n_{56}) dn_{34}dn_{56},
\end{equation}
%%%%%%%%%%%%%%%%%%%%%%%%%%%%%%%%%%%%%%%%%%%%%%%%%%%%%%%%%%%%%%%%%
where $n_{34}=n_3-n_4$, $n_{56}=n_5-n_6$ and we used the fact
that the numbers of counted photons appear in the Gaussian
exponential phase estimate (\ref{nfm_estim}) only in terms of
their differences.
The probability distribution of the differences
can be calculated from the photocount distributions
following the simple rule
%%%%%%%%%%%%%%%%%%%%%%%%%%%%%%%%%%%%%%%%%%%%%%%%%%%%%%%%%%%%%%%%%
\begin{equation}\label{joint_dif}
P(n_{ij})=\int\!\!\int P(n_i) P(n_j)
\delta(n_i-n_j-n_{ij})dn_i dn_j,
\end{equation}
%%%%%%%%%%%%%%%%%%%%%%%%%%%%%%%%%%%%%%%%%%%%%%%%%%%%%%%%%%%%%%%%%
where $ij=34,56$.
For Gaussian probability distributions we have
%%%%%%%%%%%%%%%%%%%%%%%%%%%%%%%%%%%%%%%%%%%%%%%%%%%%%%%%%%%%%%%%%
\begin{equation}\label{p_dif}
P(n_{34})=\frac{1}{\sqrt{2\pi(\bar{n}_{3}+\bar{n}_{4})}}
e^{-\frac{(n_{34}-\bar{n}_{34})^2}{2(\bar{n}_3+\bar{n}_4)}},
\end{equation}
%%%%%%%%%%%%%%%%%%%%%%%%%%%%%%%%%%%%%%%%%%%%%%%%%%%%%%%%%%%%%%%%%
for example. Since the signal to noise ratio is large in the
limit of high intensity, it is legitimate to split the
counted numbers of photons (or their differences) into
their mean values and small fluctuating parts
%%%%%%%%%%%%%%%%%%%%%%%%%%%%%%%%%%%%%%%%%%%%%%%%%%%%%%%%%%%%%%%%%
\begin{equation} \label{taylor}
n_i=\bar{n}_i+\Delta_i, \qquad i=34,56.
\end{equation}
%%%%%%%%%%%%%%%%%%%%%%%%%%%%%%%%%%%%%%%%%%%%%%%%%%%%%%%%%%%%%%%%%
Now we expand the estimated sine and cosine phase functions
keeping only the fluctuation--independent term and
the second--order terms in the fluctuations.
For $\cos\theta$ we get
%%%%%%%%%%%%%%%%%%%%%%%%%%%%%%%%%%%%%%%%%%%%%%%%%%%%%%%%%%%%%%%%%
\begin{eqnarray}\label{gauss_exp}
\frac{n_{34}}{\sqrt{n^2_{34}+n^2_{56}}}&\approx&
\frac{\bar{n}_{34}}{\sqrt{\bar{n}^2_{34}+\bar{n}^2_{56}}}-
\frac{3}{2}\frac{\bar{n}_{34}\bar{n}^2_{56}}
{(\bar{n}^2_{34}+\bar{n}^2_{56})^{5/2}}\Delta^2_{34}\nonumber\\
&+&
\frac{1}{2}\frac{\bar{n}_{34}(2\bar{n}^2_{56}-\bar{n}^2_{34})^2}
{(\bar{n}^2_{34}+\bar{n}^2_{56})^{5/2}}\Delta^2_{56}.
\end{eqnarray}
%%%%%%%%%%%%%%%%%%%%%%%%%%%%%%%%%%%%%%%%%%%%%%%%%%%%%%%%%%%%%%%%%
The expansion of $\sin\theta$ is obtained exchanging
$34\leftrightarrow 56$. Substituting Eqs. (\ref{gauss_exp})
and (\ref{p_dif}) to Eq.~(\ref{gauss_int}) and using
the following relations
%%%%%%%%%%%%%%%%%%%%%%%%%%%%%%%%%%%%%%%%%%%%%%%%%%%%%%%%%%%%%%%%%
\begin{eqnarray}\label{vztahy}
\langle\Delta^2_{34}\rangle=\bar{n}_3+\bar{n}_4, \qquad
\langle\Delta^2_{56}\rangle=\bar{n}_5+\bar{n}_6, \\
\bar{n}^2_{34}+\bar{n}^2_{56}=N^2V^2, \qquad
n_3+n_4=n_5+n_6=N,
\end{eqnarray}
%%%%%%%%%%%%%%%%%%%%%%%%%%%%%%%%%%%%%%%%%%%%%%%%%%%%%%%%%%%%%%%%%
the first set being implied by Eq.~(\ref{p_dif}), we arrive at an
approximate mean value of the cosine function of the estimated phase
%%%%%%%%%%%%%%%%%%%%%%%%%%%%%%%%%%%%%%%%%%%%%%%%%%%%%%%%%%%%%%%%%
\begin{equation}\label{gauss_rozvoj}
\langle\cos\theta\rangle\approx\cos\bar{\theta}-\frac{1}{2NV^2}
(\cos^3\bar{\theta}+\cos\bar{\theta}\sin^2\bar{\theta}).
\end{equation}
%%%%%%%%%%%%%%%%%%%%%%%%%%%%%%%%%%%%%%%%%%%%%%%%%%%%%%%%%%%%%%%%%
An analogous expression for $\sin\theta$ is obtained exchanging
$\cos\bar{\theta}\leftrightarrow\sin\bar{\theta}$ in
Eq.~(\ref{gauss_rozvoj}).
Finally, using Eq.~(\ref{gauss_rozvoj}) and a similar expression for
$\sin\theta$ in the dispersion formula (\ref{disp}) and neglecting
terms of the order $1/N^2$, we arrive at the asymptotic dispersion
of the Gaussian estimating procedure (\ref{asym_g}). Since a finite
change in the effectiveness of the estimation caused by an infinitely small
change of the estimated parameter is unphysical, the derived
expression for the asymptotic dispersion also holds for the isolated
values of the true phase for which our procedure fails.

In the case
of ML estimation we can proceed in a completely analogous way.
Starting from the expansion of the Poissonian phase estimator
(\ref{phase_p}) in fluctuations of the number of counted photons,
we obtain, by a straightforward but a rather lengthy calculation,
desired expectation values of the cosine and sine phase functions
%%%%%%%%%%%%%%%%%%%%%%%%%%%%%%%%%%%%%%%%%%%%%%%%%%%%%%%%%%%%%%%%%
\begin{eqnarray}\label{poiss_rozvoj}
\langle\cos\theta\rangle\approx\cos\bar{\theta}+
\sum_{i=3}^6C_i(\bar{\theta},V,N)\bar{n}_i.\\
\langle\sin\theta\rangle\approx\sin\bar{\theta}+
\sum_{i=3}^6S_i(\bar{\theta},V,N)\bar{n}_i.
\end{eqnarray}
%%%%%%%%%%%%%%%%%%%%%%%%%%%%%%%%%%%%%%%%%%%%%%%%%%%%%%%%%%%%%%%%%
Here $C_i$'s and $S_i$'s are the coefficients of the terms of the Taylor
series, quadratic in corresponding fluctuations $\Delta_i$. When an
explicit form of the coefficients is substituted into
(\ref{poiss_rozvoj}), it is then easy to obtain the asymptotic
dispersion (\ref{asym_p}) of the unconstrained
Poissonian ML estimator (\ref{phase_p}).

\section{Asymptotic dispersion of the
si\-ngle--parameter Poissonian estimator} \label{ap_2}

As above, we will suppose that detected numbers of photons can
be decomposed
into their mean values and fluctuating parts, small with respect
to the means, as follows, $n_i$=$\bar{n}_i+\Delta n_i$.
Inspection of the Poissonian likelihood function (\ref{lik})
shows that the point $\theta$ is a local maximum of ${\cal L}$
if and only if the condition
%%%%%%%%%%%%%%%%%%%%%%%%%%%%%%%%%%%%%%%%%%%%%%%%%%%%%%%%%%%%%%%%%
\begin{equation}\label{extrem}
\frac{d}{d\bar\theta}\ln[{\cal L}(\theta)]=0
\end{equation}
%%%%%%%%%%%%%%%%%%%%%%%%%%%%%%%%%%%%%%%%%%%%%%%%%%%%%%%%%%%%%%%%%
holds. Assuming a perfect experimental setup,
$V$=$1$, the derivative of the log--likelihood function
becomes
%%%%%%%%%%%%%%%%%%%%%%%%%%%%%%%%%%%%%%%%%%%%%%%%%%%%%%%%%%%%%%%%%
\begin{eqnarray}\label{der_lik}
{\cal L}'_{log}(\bar{\theta})
&=&-n_3
\frac{\sin\bar{\theta}}
{1+\cos\bar{\theta}}+n_4\frac{\sin\bar{\theta}}{1-\cos\bar{\theta}}
+n_5\frac{\cos\bar{\theta}}{1+\sin\bar{\theta}}\nonumber\\
&&\quad-n_6\frac{\cos\bar{\theta}}{1-\sin\bar{\theta}},
\end{eqnarray}
%%%%%%%%%%%%%%%%%%%%%%%%%%%%%%%%%%%%%%%%%%%%%%%%%%%%%%%%%%%%%%%%%
where ${\cal L}'_{log}\equiv d \ln({\cal L})/d\bar\theta$.
Now we make use of the fact that the result of unconstrained
ML estimation (\ref{phase_p}) is not so bad and in particular it
lies close to the true global maximum
of the likelihood function (\ref{lik}). The purpose is two--fold.
First, the estimated phase (\ref{phase_p}) is a good starting
point for finding a root of the expression (\ref{der_lik}) with the help
of some approximation method; second, it automatically
selects the global maximum of {$\cal L$} among all possible roots
of Eq.~(\ref{extrem}).

In order to improve our initial guess $\theta_0$:
%%%%%%%%%%%%%%%%%%%%%%%%%%%%%%%%%%%%%%%%%%%%%%%%%%%%%%%%%%%%%%%%%
\begin{equation}\label{poc_bod}
\theta_0\equiv\mbox{arg}\left\{e^{i\theta}
\right\},
\end{equation}
%%%%%%%%%%%%%%%%%%%%%%%%%%%%%%%%%%%%%%%%%%%%%%%%%%%%%%%%%%%%%%%%%
where $\exp(i\theta)$ is given by Eq. (\ref{phase_p}), we use
the Newton method. Since the second derivation of log--likelihood
always differs from zero, the algorithm converges quickly and one step
of the method is usually enough to find the global maximum with
sufficient accuracy. The improved value of the estimated phase
shift thus reads
%%%%%%%%%%%%%%%%%%%%%%%%%%%%%%%%%%%%%%%%%%%%%%%%%%%%%%%%%%%%%%%%%
\begin{equation}\label{newton}
\theta=\theta_0-\frac{{\cal L}'(\theta_0)}
{{\cal L}''(\theta_0)}.
\end{equation}
%%%%%%%%%%%%%%%%%%%%%%%%%%%%%%%%%%%%%%%%%%%%%%%%%%%%%%%%%%%%%%%%%
Now we expand the sine and cosine
phase functions around $\theta_0$ keeping terms up
to the second order in the correction.
The expectation values of the two functions become
%%%%%%%%%%%%%%%%%%%%%%%%%%%%%%%%%%%%%%%%%%%%%%%%%%%%%%%%%%%%%%%%%
\begin{eqnarray}\label{newton_cos}
E\left\{
\begin{array}{c} \cos\theta\\ \sin\theta\end{array}
\right\}&=&E\left\{
\begin{array}{c} \cos\theta_0\\ \sin\theta_0\end{array}\right\}
\pm\frac{1}{2}E\left\{
\begin{array}{c} \sin\theta_0\\ \cos\theta_0\end{array}
\frac{{\cal L}'(\theta_0)}{{\cal L}''(\theta_0)}\right\}
\nonumber\\
&&\quad-\frac{1}{6}E\left\{
\begin{array}{c} \cos\theta_0\\ \sin\theta_0\end{array}
\left[
\frac{{\cal L}'(\theta_0)}{{\cal L}''(\theta_0)}\right]^2\right\}.
\end{eqnarray}
%%%%%%%%%%%%%%%%%%%%%%%%%%%%%%%%%%%%%%%%%%%%%%%%%%%%%%%%%%%%%%%%%
Finally using Eqs.~(\ref{phase_p}), (\ref{vis_p}), (\ref{poc_bod}) and
(\ref{der_lik}) in (\ref{newton_cos}), expanding the result to
the second order in fluctuations $\Delta_i$, replacing Poissonian
photocount distributions by Gaussian with the same variance,
carrying out the average, substituting the resulting
expectation values of the sine and cosine phase functions into
Eq.~(\ref{disp}), expanding the dispersion in $1/N$ and keeping terms
at most linear in $1/N$, we arrive at the asymptotic dispersion of the
one--parameter ML phase estimation in the form
%%%%%%%%%%%%%%%%%%%%%%%%%%%%%%%%%%%%%%%%%%%%%%%%%%%%%%%%%%%%%%%%%
\begin{equation}\label{final}
\sigma^2=\frac{1}{2N}+O\left(\frac{1}{N^2}\right).
\end{equation}
%%%%%%%%%%%%%%%%%%%%%%%%%%%%%%%%%%%%%%%%%%%%%%%%%%%%%%%%%%%%%%%%%
The enormous amount of calculation work necessary to obtain this
result was carried out with the help of Maple V
symbolic mathematical language.

\end{document}